\documentclass{article}

\usepackage{arxiv}

\usepackage[utf8]{inputenc} 
\usepackage[T1]{fontenc}    
\usepackage{hyperref}       
\usepackage{url}            
\usepackage{booktabs}       
\usepackage{amsfonts}       
\usepackage{nicefrac}       
\usepackage{microtype}      
\usepackage{lipsum}
\usepackage{graphicx}
\graphicspath{ {./images/} }

\title{On the Hill and Kolmogorov statistics and the $K-$function representation
  of judiciously modified Avrami kinetics}

\author{
 Alexander M. Korsunsky \\
  MBLEM, Department of Engineering Science\\
  University of Oxford\\
  Oxford OX1 3PJ, U.K. \\
  \texttt{alexander.korsunsky@eng.ox.ac.uk} \\
  
}

\begin{document}
\maketitle
\begin{abstract}
This study of the phase transformation kinetics pertains to the processes of crystallisation of amorphous polymers, ceramics, and bulk metallic glasses (BMG), as well as solid state transformations by the nucleation and growth mechanism. It is also likely to be relevant to the description of dissolution and precipitation processes, and more generally chemical reaction kinetics, as well as biochemical processes such as ligand binding, haemoglobin oxygenation, etc. The mathematical framework of 'knee' functions is examined, and its implications are explored.
\end{abstract}


\section{Introduction}
The kinetics of reactions and solid state phase transformations is a central topic in chemistry and materials science that links the thermodynamic study of equilibria with the process of state change involving nucleation and diffusion-controlled growth of new phases. Thiis report brings together two different descriptions known as Hill and Kolmogorov-Avrami kinetics in a unified generalised approach. 

\section{Hill kinetics}

Hill kinetics arises in the context of ligand binding and was first introduced heuristically by A.V. Hill in 1910 \cite{hill}. Notably, Hill's approach was largely heuristic: he claims that he "decided to try whether equation" in the form of fractional power law expression would fit experimental observations. It is also worth noting that explicit time dependence is absent from the original Hill's expressions. 

The origins of Hill kinetics can be linked to the nature of biochemical reactions that involve reversible binding of a small molecule $L$ (called ligand) to a large macromolecule $P$,

\begin{equation}
    n L + P \leftrightarrow L_n P:
\end{equation}

Examples include oxygen binding to haemoglobin, histamine binding to histamine H2 receptor that regulates acid content in the stomach, and nutrient binding to human serum albumin \cite{ortiz}.

The extent of binding can be expressed by the proportion of bound macromolecules $[L_n P]$ in the total sum of bound and unbound macromolecules $[P] + [L_n P]$, given by the fraction

\begin{equation} \label{eq:theta}
    \theta = \frac{[L_n P]}{[P]+[L_n P]}.
\end{equation}

The fraction of {\em unbound} macromolecules is given by

\begin{equation} \label{eq:chi}
    \chi=1-\theta=\frac{[P]}{[P]+[L_n P]}.
\end{equation}

The assumption that underlies the derivation is that the rate of a chemical reaction is proportional to the product of concentrations of all the reactants. Dynamic equilibrium of concentration $[L]$ requires

\begin{equation}
    \frac{{\rm d}[L]}{{\rm d}t} = -k_b [L]^n [P] + k_u [L_n P] = 0,
\end{equation}

\begin{equation}
    [L_n P] = \frac{k_b}{k_u} [L]^n [P].
\end{equation}

Substituting this expression into equation (\ref{eq:chi}) gives

\begin{equation} \label{eq:nchi}
    \chi = \frac{1}{1+ (k_b/k_u) [L]^n }.
\end{equation}

The {\em bound} macromolecule volume fraction is given by 
\begin{equation} \label{eq:ntheta}
    \theta = \frac{1}{1+ \frac{(k_u/k_b)}{ [L]^n} }.
\end{equation}

The time dependence that is here referred to as Hill's kinetics arises if inverse time dependence is assumed for the unbound ligand concentration, $[L] \propto t^{-1}$. We write it in the form

\begin{equation}
    \frac{(k_u/k_b)}{[L]^n} = k t^n / n,
\end{equation}

that leads to the expression

\begin{equation}
    \theta = \frac{1}{1+k t^n / n}.
\end{equation}

Ortiz \cite{ortiz} remarks that while integer values of $n$ can be explained by considering chemical reaction rates, in practice non-integer $n$ values are found to provide the best fit to observed relationships, and proposes the derivation based on scale invariance. This is similar to the reasoning employed in \cite{amk2005arXiv} that uses functional equations to draw the general conclusion that self-similar processes must be described by power law relationships. For the case of Hill kinetics this implies a broader form of time dependence $[L] \propto t^{-\gamma}$, where $\gamma \neq 1$. This leads to the same form of time dependence of $\theta$ given above with the generalisation to include non-integer $n$ values.

\section{Kolmogorov-Avrami kinetics}

The so-called Avrami equation is also sometimes referred to as Johnson-Mehl-Avrami-Kolmogorov, or JMAK, in reverse chronological order of contributions from respective authors \cite{kolm},\cite{avrami}. It describes the volume fraction of transformed material as a function of time under isothermal conditions. 

We shall consider the  derivation for the case of solid state transformation from the initial phase $\alpha$ to the final phase $\beta$ at a fixed temperature, beginning at time $t=0$. The derivation is based on two key assumptions.

The first assumption is that that {\em nucleation} proceeds randomly and homogeneously over the entire untransformed part of the material at a constant rate ${\dot N}$ per time and volume. 

The second assumption is that growth proceeds uniformly over time at a constant rate of advance of the interphase boundary ${\dot G}$. This simplifying assumption ignores the effects associated with anisotropy (orientation dependence) of diffusion, interfacial energy, etc.

Instead of keeping track of the transformed volume fraction $v_\beta=V_\beta / V$ and introducing the somewhat artificial auxiliary concept of {\em extended volume} used in some derivations, we focus out attention here on the untransformed volume fraction $v_\alpha = V_\alpha / V$, noting of course that $v_\alpha + v_\beta = 1$.

Since nucleation occurs uniformly within untransformed volume, during the time interval between $\tau$ and $\tau + {\rm d}\tau$ the number of nuclei $N$ per unit sample volume is given by ${\dot N} v_\alpha {\rm d}\tau$. 

During the time between $\tau$ and $t$ each of these nuclei grows at the steady radial rate of $\dot G$ into a sphere of radius ${\dot G}(t-\tau)$ with the volume $\frac{4\pi}{3}{\dot G}^3 (t-\tau)^3$. 

At time $t$, the rate of increment of the untransformed volume fraction is given by the integral over subsidiary time $\tau$, i.e.

\begin{equation}
    \frac{{\rm d}v_\alpha}{{\rm d}t} = - v_\alpha \int_0^t \frac{4\pi}{3}{\dot G}^3 (t-\tau)^3 {\dot N} {\rm d}\tau = - v_\alpha \frac{\pi}{3}{\dot G}^3 {\dot N} t^3.
\end{equation}

This can be re-written as 

\begin{equation}
    \frac{{\rm d}v_\alpha}{{\rm d}t} =  - k v_\alpha t^4, 
\end{equation}
where $k = \frac{\pi}{3}{\dot G}^3 {\dot N}$.

Firstly, it is worth noting that the fourth power of time appearing in the volume fraction rate expression is a reflection of the system's dimensionality $D$ through $n=D+1$. The growth of nuclei in three dimensions leads to $n=3+1=4$, whilst growth of nuclei in 2D material or thin film would give rise to $n=2+1=3$. Further extension of this consideration suggests that fractal dimensions of the transformed volume may give rise to non-integral values of $n$.

We now wish to address a crucial implication embedded in this derivation, namely, the assumed proportionality of the reduction rate of the untransformed volume fraction to the volume fraction $v_\alpha$ itself. 

A number of arguments may be put forward to suggest super-linear dependence of ${{\rm d}v_\alpha}/{{\rm d}t}$ on the untransformed volume fraction $v_\alpha$, for example, the influence of the amount of solute available for the formation of $\beta$ phase present in the untransformed volume, etc. In order to allow accounting for such dependence we provide below a generalisation of the Avrami kinetics derivation to a broader range of cases.

\section{Generalisation - the {\em K-}function}

The general formulation of the rate equation for the volume fraction $v_\alpha$ of the initial phase can be written as 

\begin{equation} \label{eq:inc}
\frac{{\rm d}v_\alpha}{{\rm d}t} = - v_\alpha^{1+m} \cdot k t^{n-1}.
\end{equation}

For example, by considering the untransformed volume in multiplicative combination with the amount of solute available for the transformation within this volume, it can be reasoned that $m\geq 0$.

It is important to note that this is a general scale-invariant expression that obeys power law dependence on both the untransformed volume (unbound concentration) and time, and thus represents the most general form.

The case of strict equality $m=0$ leads to the Kolmogorov-Avrami kinetics that needs to be treated separately.

Taking first the case $m>0$, the above can be rewritten as

\begin{equation} \label{eq:incm}
{\rm d}\left(\frac{1}{v_\alpha^m}\right) = m k t^{n-1} {\rm d}t
\end{equation}

Integration gives

\begin{equation} \label{eq:int}
\frac{1}{v_\alpha^m} - 1 = \left( \frac{m}{n} \right) k t^{n} 
\end{equation}

so that finally

\begin{equation} \label{eq:kfunc}
v_\alpha (t) = \displaystyle \frac{1}{[1+(m/n) k t^{n}]^{1/m}} 
\end{equation}

This is a general form of phase transformation kinetics that we refer to as "knee function", or $K$-function. It has been introduced by the present author in a somewhat different context of material strength \cite{amk2005arXiv}. For convenience of reference we introduce the notation 

\begin{equation}
    K_{n,m}(t)= \displaystyle \frac{1}{[1+(m/n) k t^{n}]^{1/m}} 
    = \displaystyle [1+m (t/t_0)^{n}]^{-1/m}.
\end{equation}

\section{The Hill function}

The above result reduces to the Hill-Langmuir equation for the special case $m=1$:

\begin{equation} \label{eq:hill}
\displaystyle v_\alpha (t) = H_n(t) = K_{n,1}(t) =
= \frac{1}{ 1+k t^{n}/n  } 
= \frac{1}{ 1+( t/t_0)^{n} } 
\end{equation}

\section{The Avrami function}

Let us now return to the generalised transformation rate equation (\ref{eq:inc}). Considering the special case $m=0$, the rate equation assumes the form

\begin{equation} \label{eq:inclin}
\frac{{\rm d}v_\alpha}{{\rm d}t} = - v_\alpha \cdot k t^{n-1}.
\end{equation}

that integrates to

\begin{equation} \label{eq:intlin}
\log{v_\alpha} = - \frac{k t^{n}}{n} = - \left( \frac{t}{t_0} \right)^{n} 
\end{equation}

leading to the Kolmogorov-Avrami kinetics expression:

\begin{equation} \label{eq:kavrami}
v_\alpha(t) = A_n(t) = \exp{\left[- \frac{k t^{n}}{n}\right]}
= \exp{\left[- \left( \frac{t}{t_0} \right)^{n} \right]}
\end{equation}

\begin{figure}[h]
    \centering
    \includegraphics[width=12cm]{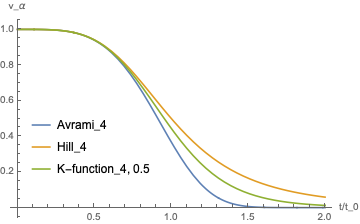}
    \caption{Comparison of K-function with Avrami and Hill functions for the case $n=4$.}
    \label{fig:ahk}
\end{figure}

\section{Limit analysis}

The unification of Hill and Avrami kinetics does not appear to be possible due to the apparent discontinuity (singularity) in the solution of the ordinary differential equation in eq. ({\ref{eq:inc}}) associated with the value $m=0$.

The underlying discontinuous dependence on the value of parameter arises already in the solution of simplest autonomous ordinary differential equation in the form

\begin{equation}
    \frac{{\rm d}y}{{\rm d}x} =  - y^{1+m}.
\end{equation}

Assuming $y(0)=1$ to be specific, the solution of this equation is written as 

\begin{equation}
     y(x) = \left\{    \begin{array}{ccr}
        (1+m x)^{-1/m}, & m \neq 0, & \\
        {\rm e}^{-x}, & (m=0). &
        \end{array}
        \right.
\end{equation}

This apparent discontinuity in the solution is resolved by the observation that the exponential solution arises by taking the limit $m \rightarrow 0+$ \footnote{A sketch proof of this equation can be obtained by setting $M=1/m$ and considering $M \rightarrow \infty$. Taking logarithm of both sides and noting for large $M$ that $-M \log(1+x/M) \cong x$, equality is obtained. \\
An alternative proof using L'Hopital's rule suggested by Bill Lionheart:  $\lim_{m \rightarrow 0} (1+m x)^{-1/m} = \exp \lim_{m \rightarrow 0} -\log(1+m x)/m = \exp(-x).$ } :

\begin{equation}
    \lim_{m \rightarrow 0} (1+ m x)^{-1/m}= {\rm e}^{-x}.
\end{equation} 
Taking the same limiting procedure for the $K-$function expression confirms that not only Hill, but also the Kolmogorov-Avrami kinetics correspond to its special cases.
Setting $m$ to progressively smaller values leads to the limiting case,

\begin{equation}
    A_n(t) = \lim_{m \rightarrow 0} K_{n,m}(t) = K_{n,0+}(t),
\end{equation}
where the notation $0+$ in the second subscript of the $K-$function is used to indicate the limiting value of this function for $m \rightarrow 0$, since formal evaluation of the function fails upon setting $m=0$.

Thus, the $K-$function represents a generalisation that provides a unified description of chemical transformation kinetics.

\section{Comparison between functions}

Comparison between the three functional expressions for the kinetics of phase transformation is illustrated in Fig.\ref{fig:ahk} for the case $n=4$. The initial behaviour of the Hill and Avrami functions is apparently closely coincident. This is confirmed by the Taylor expansions of the two functions:

\begin{eqnarray}
    H_n(t) & = & 1 - \left( \frac{t}{t_0} \right)^{n} + 
    \frac{1}{2}\left( \frac{t}{t_0} \right)^{2n} + ... \\
    A_n(t) & = & 1 - \left( \frac{t}{t_0} \right)^{n} + 
    \left( \frac{t}{t_0} \right)^{2n} + ...
\end{eqnarray}

Therefore, the Hill and Avrami kinetics lead to identical descriptions of the early stages of phase transformation process. However, these functions demonstrate fundamentally different behaviour at the later stages, particularly approaching completion of the transformation. The asymptotic approach of complete transformation is significantly faster for the case of Avrami function compared to the Hill function, as is apparent from Fig.\ref{fig:ahk}.   

As noted above, setting the value $m=1$ within the $K-$function reduces it to the Hill function. Reducing the value of $m$ causes the curve to shift towards shorter times, as illustrated in Fig.\ref{fig:ahk} for the $K-$function with the value $m=0.5$.   

\begin{figure}[h]
    \centering
    \includegraphics[width=10.5cm]{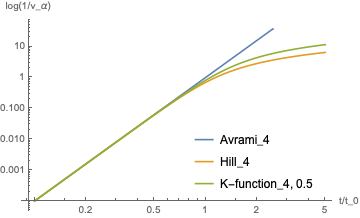}
    \caption{K-function and Avrami and Hill functions represented using double logarithmic plot.}
    \label{fig:ahklog}
\end{figure}

\begin{figure}[h]
    \centering
    \includegraphics[width=10.5cm]{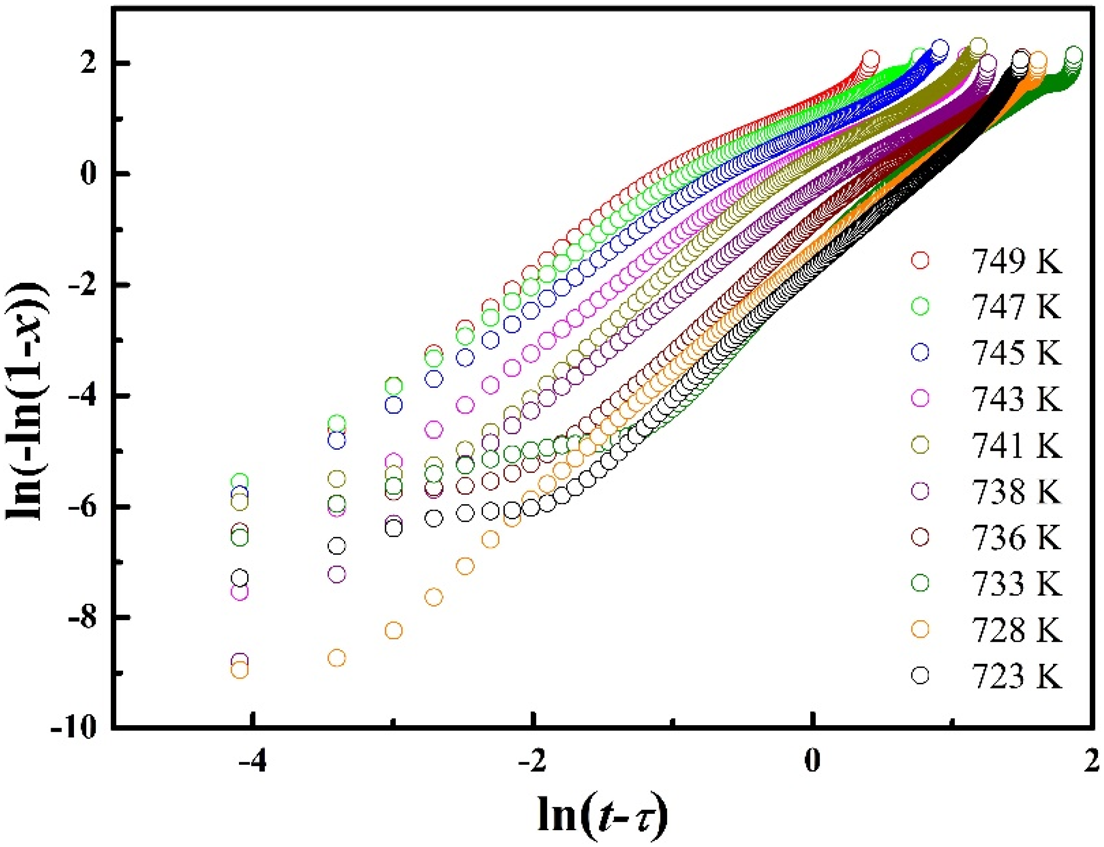}
    \caption{Crystallisation kinetics of bulk metallic glass (from S. Mandal {\em et al.} [2]).}
    \label{fig:mandal7}
\end{figure}

A popular means of representing Avrami kinetics is via the use of double logarithmic plot illustrated in Fig.\ref{fig:ahklog}.  

This way of displaying and comparing these functions highlights the deviation of Hill function (and $K-$function) at large times from the straight line corresponding to pure Avrami kinetics. In materials science, this regime concerns the important effect of transformation retardation upon nearing completion. This effect is apparent in Fig.\ref{fig:mandal7} \cite{mandal} where the retardation of crystallisation of bulk metallic glass is apparent towards long times \footnote{The final upward 'wiggle' seen at the top right end of the curves may be associated with the error of experimental data interpretation.}. In the context of the above analysis, the retardation of transformation kinetics may be associated with the transition from Avrami to Hill or $K-$function kinetics. The underlying cause of this change, as is clear from the above discussion, may be traced to the stronger than linear dependence of the transformation rate on the untransformed volume. 

\begin{figure}[h]
    \centering
    \includegraphics[width=10.5cm]{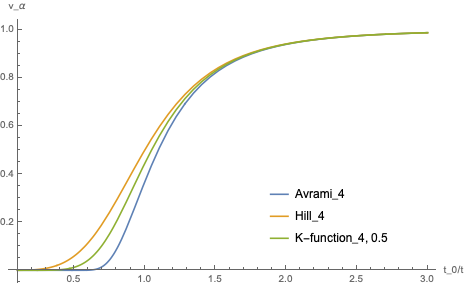}
    \caption{K-function and Avrami and Hill functions represented using double logarithmic plot.}
    \label{fig:ahk_inv}
\end{figure}

In order to investigate the nature of transformation kinetics described by the Avrami, Hill and $K-$function at large values of normalised time $t/t_0$, we generate an 'inverse time' plot with $\xi=t_0/t$ as abscissa, as shown in Fig.\ref{fig:ahk_inv} (for $n=4$). 

It can be noted once again that the different curves converge as $t_0/t \rightarrow \infty$, i.e. towards short times. It is also apparent that for long times, $t_0/t \rightarrow 0$, Avrami kinetics demonstrates the fastest convergence to zero volume fraction of untransformed phase, $v_\alpha$. Asymptotically, this behaviour is given by $A_4(\xi) \cong \exp(-1/\xi^4)$. In contrast, Hill kinetics at small values of $\xi$ (long times) is asymptotically described by the markedly slower power law function $H_4(\xi) \cong \xi^4$. The presence of the additional parameter $m$ in the definition $K-$function allows a degree of control over the rate of convergence. 

\section{Rate analysis}

\begin{figure}[h]
    \centering
    \includegraphics[width=12cm]{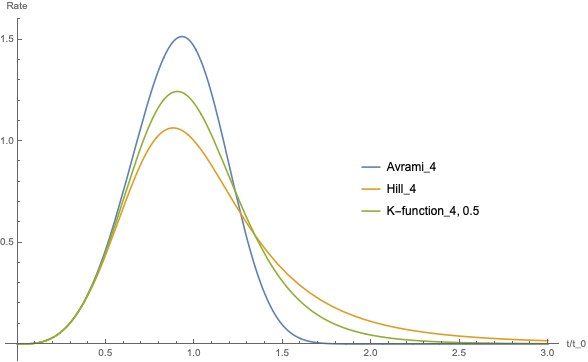}
    \caption{Comparison of transformation rate described by the derivatives of Avrami and Hill functions, and the $K-$function.}
    \label{fig:rate}
\end{figure}

The transformation rate can be obtained by differentiation of the $K-$function with respect to time, leading to
\begin{equation}
\dot{K}_{n,m}(t,t_0) = -(n/t_0) (t/t_0)^{n-1} 
\left[1 + m (t/t_0)^n\right]^{-(1+m)/m}.
\end{equation}

The transformation rate for Avrami kinetics is expressed by
\begin{equation}
\dot{A}_{n} (t,t_0) = -(n/t_0) (t/t_0)^{n-1} \exp[-(t/t_0)^n].
\end{equation}
The functional form of this rate expression is different from the Avrami function $A_n(t)$, and represents a product of exponent with a power law. It is an instance of the general exponential$\cdot$polynomial (or expolynomial) function introduced in a different context in \cite{salvati}.

Fig.\ref{fig:rate} illustrates the difference between Hill and Avrami transformation rates, and the rate decribed by the $K-$function for $n=4,\,m=0.5$.

\section{Non-isothermal kinetics}

Non-isothermal kinetics for steady cooling rate can be described by integration of the temperature-dependent transformation rate. This can be written assuming the dependence of characteristic transformation time on temperature, $t_0[T]$, with the temperature, in turn, depending on the process time, $T(t)$, the untransformed volume fraction at time $t_1$ can be expressed by the integral  

\begin{equation}
    v_\alpha(t_1) = \int_{0}^{t_1} \dot{K}_{n,m}(t,t_0[T(t)]) {\rm d}t.
\end{equation}


\section{Discussion}

The derivation of the general transformation kinetic equation presented here leads to the $K-$function expression that subsumes both Hill and Avrami kinetics functional forms traditionally applied in different scientific areas of biochemistry and materials science, respectively. It has been shown that the second parameter $m$ appearing in the generalised $K-$function form allows control over the steepness of the sigmoidal curve, and thus confers additional flexibility for precise process description and deeper interpretation of the underlying physics of specific phenomena.

\bibliographystyle{unsrt}  


\end{document}